# Calculation of Optical Response Functions of Dilute-N GaPAsN Lattice-matched to Si

Y. Zou* and S. M. Goodnick

*School of Electrical, Computer and Energy Engineering, Arizona State University,*

*Tempe, Arizona, 85287, USA*

*Yongjie.Zou@asu.edu

Dilute-N GaPAsN alloys have great potential for optoelectronics lattice-matched to Si. However, there is a lack of systematic calculation of the optical response of these alloys. The present paper uses the $sp^3d^5s^*s_N$ tight-binding model to calculate the fullband electronic structure of dilute-N GaPAsN, and then calculate the optical response functions considering direct transitions within the electric dipole approximation. Good agreement is obtained for the dielectric function in comparison to available optical data for dilute nitrides. To achieve this, the $sp^3d^5s^*$ parameters for GaP and GaAs are optimized for their optical properties in comparison to published data, which are then used as the basis for the $sp^3d^5s^*s_N$ parameters for dilute-N GaPN and GaAsN. The calculated absorption between the valence band and the newly formed lowest conduction band of the dilute nitrides increases as the N fraction increases, in agreement with experiments, mainly due to the net increase in their coupling in the entire Brillouin zone, supported by the calculated momentum matrix element in the present work.

## I. INTRODUCTION

The idea of direct bandgap materials grown on Si is very attractive in the optoelectronic community, as Si is the low-cost foundation for the semiconductor industry. However, to ensure good device performance, lattice-matched structures are needed to avoid the formation of performance degrading defects such as threading dislocations. GaP$_{1-x-y}$As$_y$N$_x$ alloys with $x < 0.05$ have shown the ability to be grown lattice-matched to Si[1,2] when the chemical composition is equivalent to (GaP$_{0.979}$N$_{0.021}$)$_x$(GaAs$_{0.805}$N$_{0.195}$)$_{1-x}$, and have a direct bandgap that can be varied by roughly 0.5 eV in the optical energy range[3–5]. With these features combined, dilute-N GaPAsN alloys have great potential for applications in lasers/light emitting diodes on Si[6–8], optoelectronic integrated circuits[1,9,10], and multijunction photovoltaics on Si[3,11–15].

For modeling dilute-N GaPAsN alloys for different applications, it is desirable to generate their optical response functions over a large photon energy range for arbitrary compositions. There are a few reports on the experimentally measured absorption and dielectric functions for a small sample of the entire quaternary system[4,5,16–25], and most of the absorption data are near the bandgap only. To the best of the authors' knowledge, there is a lack of publication on the direct calculation of the optical response functions of dilute-N GaPAsN alloys, such as their refractive indices, absorption coefficients, or dielectric functions, etc. Benkabou *et al.*[26] used a virtual crystal approximation with the empirical pseudopotential method to calculate the bandgaps of GaPN alloys, and based on these bandgaps, calculated the refractive indices for GaPN using a closed-form model by Herve and Vandamme[27]. One should note that they predicted an indirect-to-direct crossover near [N]=0.27, which is incorrect as explained later, and the refractive indices are limited to only the static region. Perlin and coworkers[28,29] calculated the absorption coefficient of $Ga_{0.96}In_{0.04}As_{0.99}N_{0.01}$ based on an estimation of the momentum matrix element with the band anticrossing (BAC) model[30]. However, their calculation limited the conduction band contributions to only $E_-$ and $E_+$ at $\Gamma$, and required empirical scaling. Robert *et al.*[31] used the tight-binding method[32,33] to calculate the band structure and optical gain of dilute-N GaPAsN/GaPN quantum wells with biaxial strain on silicon substrates, but there is not much discussion on their optical properties. Laref *et al.*[34] calculated the optical functions of hexagonally structured GaPN using a non-relativistic full-potential linearized augmented plane wave (FP-LAPW) method in the framework of the density functional theory (DFT). The calculation only included [N]=0.25, 0.50, 0.75 as the alloy fractions, probably to avoid computational burden, which lies outside the dilute nitrogen range. Recently, Polak *et al.*[35] used the unfolded band structure calculated with the DFT method and a 128-atom supercell to investigate dilute-N GaPAsN on the possibility of the formation of an intermediate band and carrier localization. Their calculated energy gaps for the binary III-Vs are quite accurate, but are low for the dilute-N alloys. In the present work, we demonstrate a computationally efficient atomistic approach to calculate the dielectric functions and absorption coefficients of dilute-N GaPAsN alloys that are lattice matched to Si, and compare with available experimental data for this system.

## II. CALCULATION METHOD

The calculation of optical functions requires proper modeling of the electronic structure, which is quite different from "normal" alloys for dilute-N GaPAsN. Their extraordinary bandgap tunability is common among a larger group of alloys called dilute nitrides (DNs), for their low nitrogen concentration, or highly mismatched alloys (HMAs), for the large differences in the atomic sizes and electronegativity of the component group-V elements. The tunability of their bandgaps is due to the large bandgap reduction caused by the incorporation of a small fraction of N atoms. Accurate description of the bandgap behaviors becomes a critical part of modeling the DNs. Baillargeon *et al.*[36] modeled the bandgap bowing for GaPN using the dielectric theory of electronegativity[37], and predicted that $GaP_{1-x}N_x$ becomes metallic when $0.3 < x < 0.6$, although they noted a possibility of a miscibility gap preventing formation of an alloy in that region. They also predicted an indirect-to-direct crossover occurring near $x = 0.46$. This contradicts the later measurements that show DN GaPN has a direct bandgap for $x$ as small as $0.0043$[5,20,38–40]. A crossover at $x = 0.03$ was calculated using an empirical pseudopotential method with 512-atom supercells[41], still in disagreement with the experiment just mentioned. Shan *et al.*[38] explained this direct nature of the bandgap in DN GaPN in terms of coupling between the extended Γ states of the host material and the localized nitrogen states, expressed as the band anticrossing (BAC) model, which was first introduced to explain the bandgap phenomena of DN GaInAsN[30]. Although the two-level BAC model ignores the details of different N states (isolated N atoms, N neighbors, N clusters, etc.[42–45]), its effectiveness in terms of experimental fit and simplicity make it useful for quick calculation of bandgaps and effective masses. For accurate optical calculation, the electronic structure over the entire Brillouin zone (fullband structure) should be accounted for. Using the BAC model to do this requires the input of the fullband structure of the host material, and coupling constants at all $\boldsymbol{k}$ points, which is possible but is not typically done considering the tradeoff between accuracy gain and increase of workload with this method. The $sp^3d^5s^*$ tight-binding model[46] is widely used for fullband calculations of semiconductors, due to the balance between accuracy and computational complexity. Following the BAC, Shtinkov et al.[33] added an $s_N$ orbital ($E_s^N = 1.725\ eV$) to a low-temperature $sp^3d^5s^*$ tight-binding model of GaAs to simulate the localized nitrogen states, and a hopping integral $s_c s_N \sigma = -1.00\sqrt{x}\ eV$ to simulate the coupling between the localized states and the extended states. By doing this, they were able to reproduce the large bandgap reduction of DN GaAsN at Γ, but also the less perturbed conduction band at $L$ and $X$ (the latter being almost unaffected). Following this, Turcotte et al.[47] added $E_s^N = 2.15\ eV$ and $s_c s_N \sigma = -1.02\sqrt{x}\ eV$ to a low-temperature $sp^3d^5s^*$ model of GaP to calculate the electronic structure of DN GaPN at 0 K. Robert et al.[31] used $E_s^N = 1.65\ eV$ and $s_c s_N \sigma = -1.04\sqrt{x}\ eV$ for DN GaAsN and $E_s^N = 2.19\ eV$ and $s_c s_N \sigma = -1.09\sqrt{x}\ eV$ for DN GaPN with a linear interpolation and a distanced law to calculated the low-temperature electronic structure of strained DN GaPAsN.

In this work, we adopt the Shtinkov model ($sp^3d^5s^*s_N$) to calculate the electronic structure and optical response of unstrained DN GaPAsN at room-temperature for lattice-matched applications on Si. The currently available room-temperature $sp^3d^5s^*$ parameters for GaP and GaAs are not optimized for their optical properties. Thus, we first optimize the $sp^3d^5s^*$ parameters for GaP and GaAs to fit their room-temperature optical properties (Section III). We then obtain the extra parameters ($E_s^N$ and $s_c s_N \sigma$) of the $sp^3d^5s^*s_N$ model for DN GaPN and GaAsN by fitting to measured bandgaps, and compare the calculated optical functions with experiment (Section IV). Finally, we use a linear interpolation to obtain the $sp^3d^5s^*s_N$ parameters for DN GaPAsN and calculate their optical response (Section V). For the optical calculations throughout this work, the imaginary part of the complex dielectric function $\varepsilon(\omega) = \varepsilon_1(\omega) + i\varepsilon_2(\omega)$ is given by the electric dipole approximation as[48]

$$\varepsilon_2(\omega) = \frac{1}{4\pi\epsilon_0}\left(\frac{2\pi e}{m\omega}\right)^2 \sum_{\mathbf{k}} |P_{cv}|^2 \delta(E_c(\mathbf{k}) - E_v(\mathbf{k}) - \hbar\omega), \qquad (1)$$

where $\omega$ is the angular frequency of the photon, $\mathbf{k}$ is the wave vector in reciprocal space, $P_{cv} = \langle c|\hat{\mathbf{e}} \cdot \mathbf{p}|v\rangle$ is the momentum matrix element, $\hat{\mathbf{e}}$ is a unit vector parallel to the electric field of the electromagnetic wave, and $\mathbf{p}$ is the momentum operator. Indirect transitions are ignored here. Since $\varepsilon$ is a linear response function, once we obtain $\varepsilon_2$, $\varepsilon_1$ is given by the Kramers-Kronig relation

$$\varepsilon_1(\omega) = 1 + \frac{2}{\pi}\mathcal{P}\int_0^\infty \frac{\omega' \varepsilon_2(\omega') d\omega'}{\omega'^2 - \omega^2}, \qquad (2)$$

where $\mathcal{P}$ denotes the Cauchy principle value of the integral. Assuming the relative permeability equals to 1, the refractive index $\tilde{n} = n + ik = \sqrt{\varepsilon}$,

$$n(\omega) = \sqrt{\frac{|\varepsilon(\omega)| + \varepsilon_1(\omega)}{2}}, \tag{3a}$$

$$k(\omega) = \sqrt{\frac{|\varepsilon(\omega)| - \varepsilon_1(\omega)}{2}}. \tag{3b}$$

The absorption coefficient is related to $k$ by

$$\alpha(\omega) = \frac{4\pi k(\omega)}{\lambda_0}. \tag{4}$$

where $\lambda_0$ is the wavelength of the light in vacuum. The momentum operator in (1) is obtained from the tight-binding (TB) Hamiltonian $H$ as[49]

$$\boldsymbol{p}(\boldsymbol{k}) = \frac{m_0}{\hbar} \nabla_{\boldsymbol{k}} H(\boldsymbol{k}), \tag{5}$$

where $H(\boldsymbol{k})$ is the TB Hamiltonian. The TB Hamiltonian matrix elements are given by[32]

$$H_{ij}^{mn} = \sum_{\boldsymbol{R}_j} e^{i\boldsymbol{k}\cdot(\boldsymbol{R}_j - \boldsymbol{R}_i)} \int \psi_n^*(\boldsymbol{r} - \boldsymbol{R}_i) H \psi_m(\boldsymbol{r} - \boldsymbol{R}_j) d^3\boldsymbol{r}, \tag{6}$$

where $\boldsymbol{R}_i$ and $\boldsymbol{R}_j$ are the positions of the basis atoms $i$ and $j$ on which the orbitals $\psi_n$ and $\psi_m$ are located, respectively. The integral in (6) can be expressed as a sum of on-site energies ($j = i, m = n$) and Slater-Koster-type hopping integrals ($j \neq i$). The exponential factors of the on-site energies become unity, thus (5) is left with a derivative of the hopping integrals.

## III. OPTIMIZATION OF THE $sp^3d^5s^*$ PARAMETERS FOR THE OPTICAL PROPERTIES OF GaP and GaAs

The published $sp^3d^5s^*$ parametrizations[46,50] are typically fit to the relevant band edge energies and effective masses, which is good for electrical calculations. For optical applications, one should also include optical properties in the optimization. Here we fit the $sp^3d^5s^*$ model to match not only the typical band edge energies and effective masses, but also critical optical transition energies.

For a zincblende III-V material, the $sp^3d^5s^*$ model comprises 31 independent parameters, which include 8 on-site energies, 21 two-center hopping integrals, and 2 spin-orbit interaction energies[46]. Fitting a set of 31 parameters is a non-trivial global optimization problem. A genetic algorithm (GA) mimics the natural selection process to obtain high-quality solutions to an optimization problem. Deaven and Ho[51] found that GA outperforms simulated annealing in molecular geometry optimization for fullerene structures. Klimeck *et al*.[52] used a GA to obtain the nearest-neighbor $sp^3s^*$ parameters (9-dimensional) and second nearest-neighbor $sp^3s^*$ parameters (20-dimensional) for Si. Here, to solve the 31-dimensional nonlinear global optimization problem, we pair our in-house TB solver with an open-source parallel genetic algorithm library, PGAPack[53], developed by Argonne National Laboratory.

At the beginning, the 31 parameters are randomly initialized as "genes" within reasonable parameter boundaries and are then packaged together into a "chromosome". According to the specified population size, multiple chromosomes are randomly created to form the initial generation. During each generation, each chromosome is evaluated by calling the $sp^3d^5s^*$ TB solver, generating relevant band parameters, and comparing them to the target values with specific weights. The "chromosome" that generates band parameters closest to the target values is given the highest rank, and so on. A selected number of the highest-ranked chromosomes will survive the generation, and give birth to the next generation through crossover, and mutation. Thus, the next generation goes through the same random process. All generations have the same population size. One can assume a good solution has been reached when a particular number of generations have passed, when the last few generations are extremely similar, or when almost all the chromosomes are the same in the generation. Factors that may change the fit of the solution include (a) the size of the population, (b) ending criteria, (c) weights of the target, (d) survival rate, (e) probabilities of crossover and mutation, and (f) the seed value for the random number generator. For the same number of evaluations, which equals to the product of population size and number of iterations, our experience is that choosing a larger population size, with respect to a larger number of iterations, usually gives better results, similar to a conclusion from using GA for information retrieval[54]. Some useful discussions on the usage of the PGAPack can be found elsewhere[55]. The fitting procedure is summarized in Figure 1.

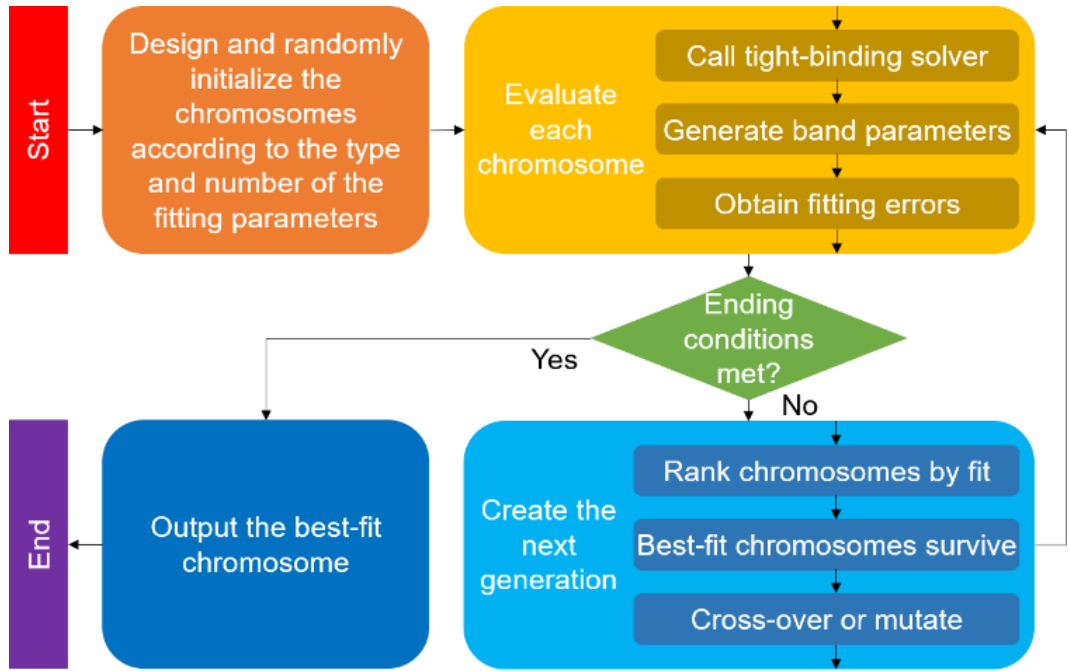

FIG. 1. Flow chart of the fitting procedure for the tight-binding parameters.

We carried out the optimization procedure for GaP and GaAs to fit experimental band energies for the conduction band bottom at $\Gamma, L, X$, the valence band split-off energy, and the effective masses at these points, as well as optical transition energies, $E_1$, $E_1 + \Delta_1$, $E_0'$, and $E_2$. These optical transitions are marked in the electronic structure of GaAs in Figure 2. For the effective masses that have relatively large uncertainties, we put less weight on fitting these values. After the fitting procedure above, we obtained the Slater-Koster-type $sp^3d^5s^*$ parameters for GaP and GaAs, which are listed in Table I.

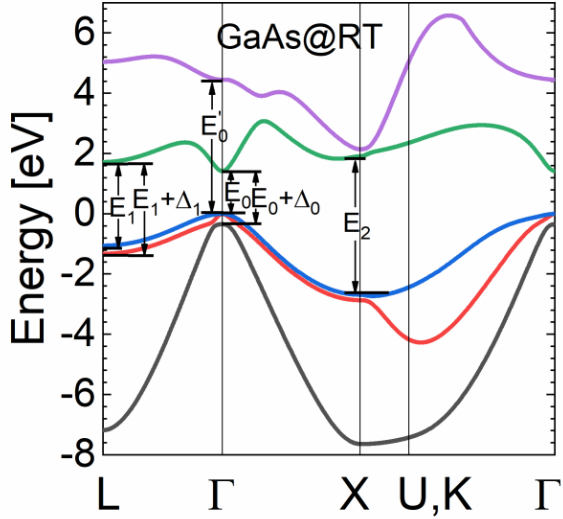

FIG. 2. Optical transition energies $E_0$, $E_0 + \Delta_0$, $E_1$, $E_1 + \Delta_1$, $E_0'$, and $E_2$ for GaAs.

TABLE I. The optimized Slater-Koster-type nearest-neighbor $sp^3d^5s^*$ parameters for GaP and GaAs at room temperature. The lattice constants are in units of Å, and all other parameters are in units of eV. The parameter notation is consistent with that of Jancu et al.'s[46].

| Parameters | GaP | GaAs |
|---|---|---|
| $a_0$ | 5.4508 | 5.6532 |
| $E_s^c$ | -0.71823 | -0.36374 |
| $E_p^c$ | 6.21668 | 8.13245 |
| $E_d^c$ | 12.85561 | 20.62765 |
| $E_{s^*}^c$ | 19.63221 | 14.12705 |
| $E_s^a$ | -6.07142 | -9.66064 |
| $E_p^a$ | 2.07966 | 2.48628 |
| $E_d^a$ | 15.03221 | 21.71325 |
| $E_{s^*}^a$ | 20.26761 | 15.18635 |
| $ss\sigma$ | -1.85072 | -1.85238 |
| $s_c p_a \sigma$ | 3.37883 | 2.82860 |
| $s_c d_a \sigma$ | -2.38353 | -2.33514 |
| $s_c s_a^* \sigma$ | -1.16664 | -1.48694 |
| $s_a p_c \sigma$ | 2.18465 | 2.50423 |
| $s_a d_c \sigma$ | -3.11886 | -1.24536 |
| $s_a s_c^* \sigma$ | -0.66599 | -0.59257 |
| $pp\sigma$ | 3.62258 | 4.04989 |
| $pp\pi$ | -0.99926 | -1.60088 |
| $p_c d_a \sigma$ | -1.86188 | -1.89807 |
| $p_c d_a \pi$ | 1.96330 | 3.82055 |
| $p_c s_a^* \sigma$ | 2.49661 | 1.89681 |
| $p_a d_c \sigma$ | -0.89955 | -2.30529 |
| $p_a d_c \pi$ | 1.25412 | 2.36166 |
| $p_a s_c^* \sigma$ | 3.71479 | 3.16348 |
| $dd\sigma$ | -0.87592 | -3.07285 |
| $dd\pi$ | 2.12065 | 4.76324 |
| $dd\delta$ | -1.51045 | -1.34699 |
| $d_c s_a^* \sigma$ | -0.07141 | -0.34281 |
| $d_a s_c^* \sigma$ | -0.25448 | -0.29833 |
| $s^* s^* \sigma$ | -4.02230 | -2.16846 |
| $\Delta_c/3$ | 0.00002 | 0.00551 |
| $\Delta_a/3$ | 0.01967 | 0.13714 |

Taking the $sp^3d^5s^*$ parameters in Table I, the TB solver generated the electronic structure of GaAs and GaP at room temperature, which are plotted along high-symmetry lines of the first Brillouin zones of these three face-centered cubic lattices in Figure 3. These plots have been shifted so that their valence band maxima are all at 0 eV. Given the availability of literature reports, the electronic structure of GaAs generated from our optimized $sp^3d^5s^*$ parameters is compared to that from quasiparticle self-consistent GW (QSGW) theory[56], as shown in Figure 3. The QSGW calculation was carried out for low temperature, and we shifted their conduction bands -0.1 eV to approximate the temperature effect. The QSGW calculation does not account for spin-orbit coupling, and we shifted the second top-most valence band from the QSGW calculation by -0.34 eV for comparison. Except for the systematically overestimated gaps of the QSGW structure[56], the two have general agreements in shape, especially for the lowest conduction band between L and Γ and for the split-off valence band.

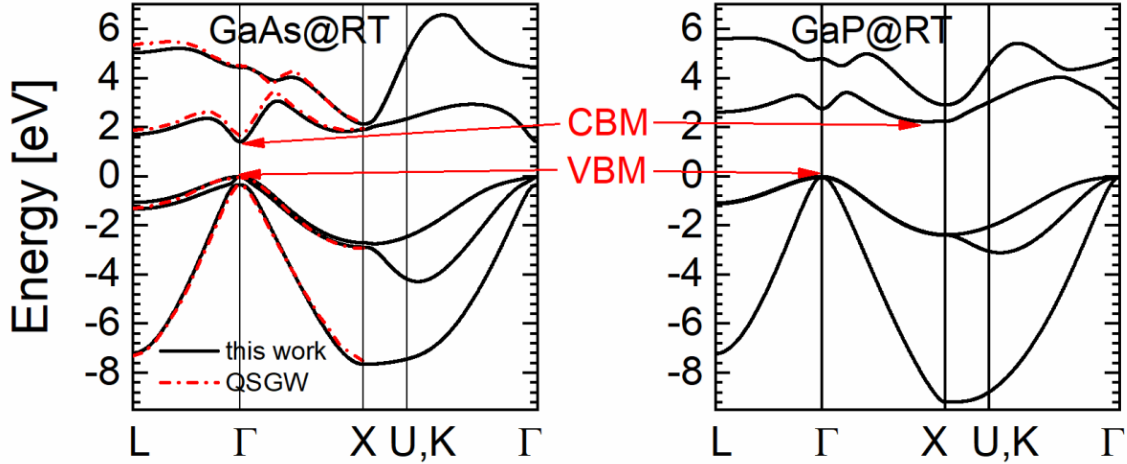

FIG. 3. Energy dispersion along high-symmetry lines for GaAs and GaP at room temperature (black solid lines). The energy dispersion of GaAs by this work is compared to that from a quasiparticle self-consistent GW calculation[56] (red dash-dot lines). The arrows indicate the conduction band minimum (CBM) and valence band maximum (VBM).

We then compare the relevant band energies, effective masses, and optical transition energies from the present models with the target values in Tables II and III. The first four columns of these tables are values from experiments and two notable compilations, and they are used as a reference to evaluate the calculated results. All the TB calculations included in the tables are based on the nearest-neighbor $sp^3d^5s^*$ model. For GaP, the available calculations of band parameters are not as abundant as those for GaAs. We used a TB parameter set published for low temperature[46], and applied hydrostatic tensile strain, from lattice constant $a(0K)$ to $a(300K)$, for comparison. Our TB results show good agreement with the reference, especially for the conduction band energies, $E_1$ and $E_2$, though the split-off energy is overestimated and the hole effective masses have larger errors than those from the two previous TB works. For GaAs, TB can generally generate conduction band energies much closer to experimental values than first-principles calculations, e.g. the QSGW calculation mentioned above, and a local-density approximation (LDA) with self-energy corrections[57], given the empirical nature of the TB calculations. Our TB results have similar effective masses and conduction band energies as two previous TB works[58,59], but have improved optical transition energies, as the latter is one of the main focuses of our fitting procedures. Overall, our optimized TB parameter sets for GaP and GaAs give reasonably good agreement to commonly referred band values, including optical transition energies.

TABLE II. Comparison of room-temperature band parameters of GaP from experiment and different calculations. $E_\Gamma$, $E_X$, and $E_L$ are energies w.r.t. the top of the valence band, and other energies are gap energies. All energies are in units of eV, and relative effective masses are unitless.

| Parameter | Exp. A[a] | Exp. B[b] | Cmp. A[c] | Cmp. B[d] | TB I[e] | TB II[f] | LDA[g] | This work |
|---|---|---|---|---|---|---|---|---|
| $E_\Gamma$ | 2.76 | | 2.92 | 2.76 | 2.85 | 2.79 | 2.05 | 2.7662 |
| $E_X$ | | | 2.27 | 2.26 | 2.36 | 2.31 | 1.80 | 2.2536 |
| $E_L$ | | | 2.64 | 2.63 | 2.55 | 2.53 | | 2.6249 |
| $E_0'$ | 4.78 | 4.8 | | 4.74 | 4.46 | 4.62 | | 4.79 |
| $\Delta_0$ | | | 0.08 | 0.085 | 0.080 | 0.090 | | 0.041 |
| $E_1$ | 3.7 | 3.73 | | 3.71 | 3.64 | 3.92 | 3.52 | 3.70 |
| $\Delta_1$ | < 0.1 | | | 0.055 | 0.1 | 0.09 | | 0.035 |
| $E_2$ | 5.05 | 5.08 | | 5.28 | 5.40 | 5.71 | ~4.65 | 5.10 |
| $m^*_{e,\Gamma}$ | | | 0.13 | 0.114 | 0.127 | 0.13 | | 0.114 |
| $m^*_{e,Xl}$ | | | 2.0 | 6.9 | 15.3 | 0.82 | | 3.08 |
| $m^*_{e,Xt}$ | | | 0.253 | 0.252 | 0.26 | 0.21 | | 0.31 |
| $m^*_{e,Ll}$ | | | 1.2 | 1.18 | 1.77 | 1.59 | | 3.42 |
| $m^*_{e,Lt}$ | | | 0.15 | 0.15 | 0.39 | 0.42 | | 0.58 |
| $m^*_{hh,[100]}$ | | | | 0.34 | 0.35 | 0.38 | | 0.488 |
| $m^*_{hh,[110]}$ | | | | 0.53 | 0.72 | 0.71 | | 0.841 |
| $m^*_{hh,[111]}$ | | | | 0.66 | 0.98 | 0.91 | | 1.060 |
| $m^*_{lh,[100]}$ | | | | 0.20 | 0.15 | 0.16 | | 0.124 |
| $m^*_{lh,[110]}$ | | | | 0.16 | 0.12 | 0.13 | | 0.112 |
| $m^*_{lh,[111]}$ | | | | 0.15 | 0.09 | 0.10 | | 0.084 |
| $m^*_{so}$ | | | 0.25 | 0.34 | 0.22 | 0.46 | | 0.200 |

[a]Extracted from the optical spectra measured by Aspnes and Studna (1983)[60].
[b]Extracted from the optical spectra measured by Jellison (1992)[61].
[c]Compilation from Vurgaftman, Meyer, and Ram-Mohan (2001)[62].
[d]Compilation from Adachi (2005)[63].
[e]From tight-binding calculation with tensile hydrostatic strain to Jancu *et al.*'s model for 0K[46].
[f]From tight-binding calculation using NEMO5[50].
[g]From LDA with lifetime broadening and self-energy corrections by Wang and Klein. (1981)[57].

TABLE III. Comparison of room-temperature band parameters of GaAs from experiment and different calculations. $E_\Gamma$, $E_X$, and $E_L$ are energies w.r.t. the top of the valence band, and other energies are gap energies. All energies are in units of eV, and relative effective masses are unitless.

| Parameter | Exp. A[a] | Exp. B[b] | Cmp. A[c] | Cmp. B[d] | TB I[e] | TB II[f] | LDA[g] | QSGW[h] | This work |
|---|---|---|---|---|---|---|---|---|---|
| $E_\Gamma$ | | 1.42 | 1.42 | 1.43 | 1.4159 | 1.416 | 1.21 | 1.67 | 1.4188 |
| $E_X$ | | | 1.90 | 1.91 | 1.9015 | 1.910 | | 1.95 | 1.9099 |
| $E_L$ | | | 1.71 | 1.72 | 1.7012 | 1.708 | | 1.89 | 1.7207 |
| $E'_0$ | | 4.44 | | ~4.54 | | | | 4.53 | 4.45 |
| $\Delta_0$ | | 0.37 | 0.34 | 0.341 | 0.3265 | 0.367 | | | 0.34 |
| $E_1$ | 2.91 | 2.91 | | ~2.9 | 2.76 | 3.28 | 2.90 | 3.18 | 2.89 |
| $\Delta_1$ | 0.18 | 0.22 | | 0.222 | 0.27 | 0.31 | | | 0.24 |
| $E_2$ | 4.78 | 4.96 | | ~5.2 | 4.34 | 5.47 | ~4.4 | 4.85 | 4.81 |
| $m^*_{e,\Gamma}$ | | | 0.067 | 0.067 | 0.0657 | 0.067 | | 0.077 | 0.067 |
| $m^*_{e,Xl}$ | | | 1.3 | 1.3 | 1.8808 | 1.480 | | | 0.36 |
| $m^*_{e,Xt}$ | | | 0.23 | 0.23 | 0.1753 | 0.204 | | | 0.16 |
| $m^*_{e,Ll}$ | | | 1.9 | 1.9 | 1.7275 | 1.446 | | | 1.65 |
| $m^*_{e,Lt}$ | | | 0.075 | 0.075 | 0.0967 | 0.136 | | | 0.36 |
| $m^*_{hh,[100]}$ | | | 0.35 | 0.33 | 0.3769 | 0.337 | | | 0.328 |
| $m^*_{hh,[110]}$ | | | 0.64 | 0.58 | 0.6566 | 0.619 | | | 0.600 |
| $m^*_{hh,[111]}$ | | | 0.89 | 0.78 | 0.8391 | 0.813 | | | 0.786 |
| $m^*_{lh,[100]}$ | | | 0.090 | 0.090 | 0.0825 | 0.083 | | | 0.082 |
| $m^*_{lh,[110]}$ | | | 0.081 | 0.080 | 0.0755 | 0.074 | | | 0.074 |
| $m^*_{lh,[111]}$ | | | 0.078 | 0.077 | 0.0736 | 0.072 | | | 0.055 |
| $m^*_{so}$ | | | 0.172 | 0.165 | 0.1624 | 0.160 | | | 0.156 |

[a]Extracted from the optical spectra measured by Aspnes and Studna (1983)[60].
[b]From experimental data in Lautenschlager *et al.* (1987)[64].
[c]Compilation from Vurgaftman, Meyer, and Ram-Mohan (2001)[62].
[d]Compilation from Adachi (2005)[63].
[e]From tight-binding calculation by Boykin *et al.* (2002)[58].
[f]From tight-binding calculation by Tan *et al.* (2015)[59].
[g]From LDA with lifetime broadening and self-energy corrections by Wang and Klein. (1981)[57].
[h]Extracted from the QSGW energy bands[56] with the conduction bands shifted -0.1 eV.

Figures 4 and 5 show comparisons of the imaginary part of the dielectric functions and absorption coefficients of GaP and GaAs calculated from our TB parameter sets, and those from experiments. To demonstrate the advantageous outcome of the optimization we performed with optical fitting targets, we have also included in the comparison the results calculated from other widely used $sp^3d^5s^*$ TB parametrizations[46,50,58]. Jancu's parametrization[46] is given at low temperature. The calculations shown here strains the lattice to its room-temperature spacing to account for the non-zero temperature effects. The "NEMO5(H)" parameters refer to a set in the NEMO5[50] input file that are mapped[59] to results from the *ab initio* HSE06 hybrid functional method. Some relatively well-matched $\varepsilon_2$ functions calculated with *ab initio* methods[57,65] are also included for comparison. The optical functions calculated from our TB parameterization show excellent agreement with the experimental data in terms of the absorption edge, the $E_1$ and $E_2$ peak positions, and the spin-orbit interaction peak positions for both III-V materials, while experimental peak widths are somewhat broader as we did not include lifetime broadening or indirect transitions. The feature positions and shapes are indicative that the present TB parameter sets should generate more accurate optical properties than other $sp^3d^5s^*$ TB parameter sets in comparison.

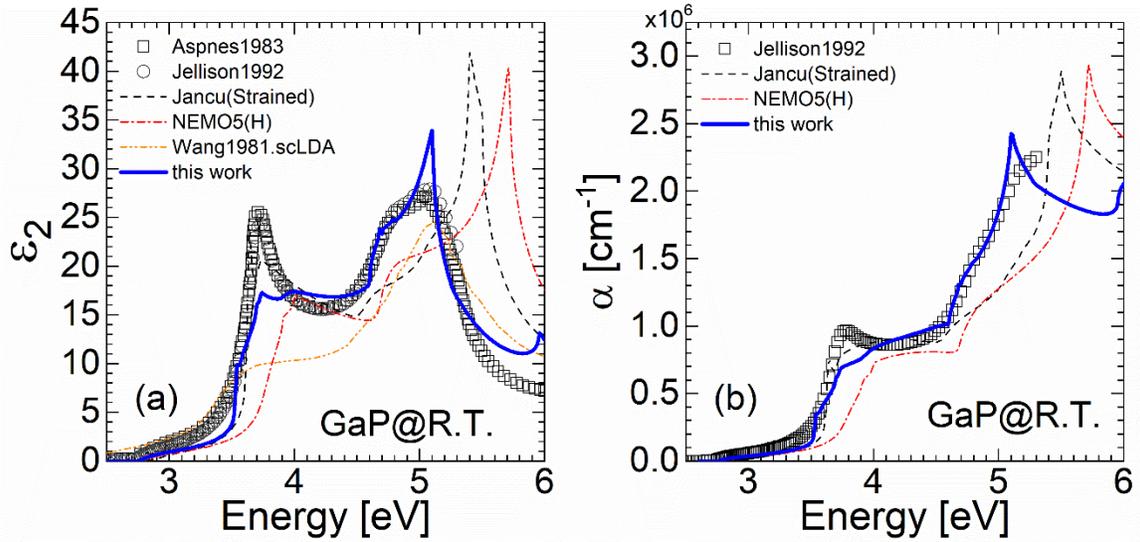

FIG. 4. The imaginary part of the dielectric function and absorption coefficient of GaP from experiment (open symbols)[61], this work (solid blue curves), and other calculations (broken curves)[46,50,57].

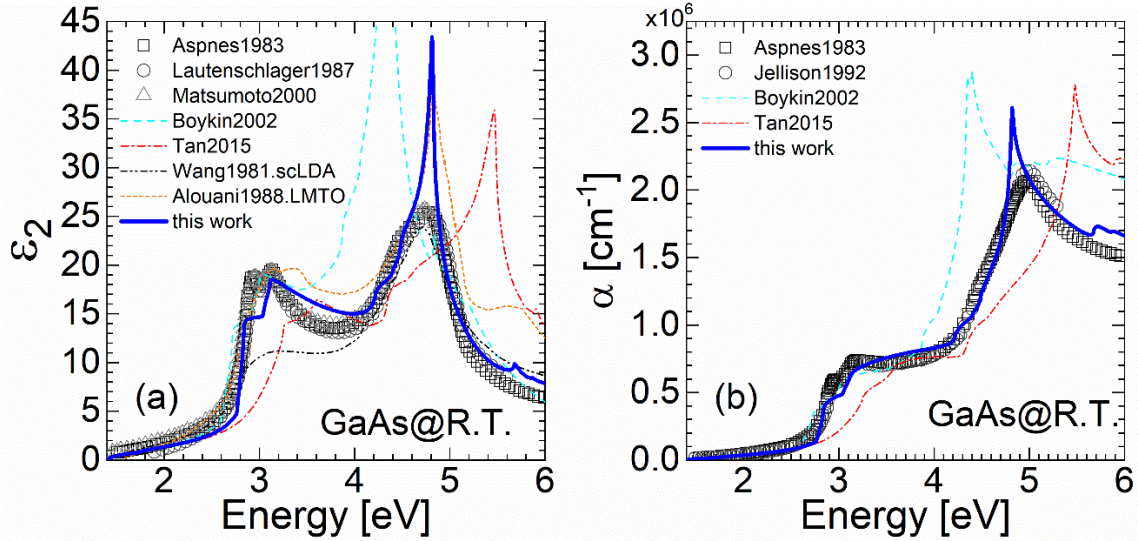

FIG. 5. The imaginary part of the dielectric function and absorption coefficient of GaAs from experiment (open symbols)[22,60,61,64], this work (solid blue curves), and other calculations (broken curves)[57–59,65].

## IV. $sp^3d^5s^*s_N$ MODELS FOR DILUTE-N GaPN and GaAsN

As shown in the previous section, the $sp^3d^5s^*$ parametrization we optimized for GaP and GaAs provides a good starting point for obtaining the $sp^3d^5s^*s_N$ parameters for DN GaPN and GaAsN for optical calculations. Since the $sp^3d^5s^*s_N$ model for AB$_{1-x}$N$_x$ is just the $sp^3d^5s^*$ model for AB plus a perturbation, implemented through $E_s^N$ and $s_cs_N\sigma$, one can model the DNs by determining the two extra parameters without changing the 31 $sp^3d^5s^*$ parameters. The $E_s^N$ and $s_cs_N\sigma$ for DN GaPN and GaAsN are determined by fitting to experimental bandgaps, and are summarized in Table IV.

TABLE IV. The on-site energies, $E_{s_N}$ and the coupling terms, $s_cs_N\sigma$ used in the $sp^3d^5s^*s_N$ calculations for GaPN and GaAsN at room temperature. Units are in eV.

|  | $E_{s_N}$ | $s_cs_N\sigma$ |
|---|---|---|
| GaP$_{1-x}$N$_x$ | 2.18 | $-1.2\sqrt{x}$ |
| GaAs$_{1-x}$N$_x$ | 1.65 | $-0.93\sqrt{x}$ |

Figure 6a shows the interacting N impurity level and GaP conduction band states transform into the new conduction bands of DN GaPN. The N effect is strongest at the zone center (Γ), becomes weaker towards $L$, and almost diminishes towards $X$; The higher conduction bands and the valence bands are mostly intact. Figure 6b marks the important transition energies in DN GaPN. The $E_{0+}$, $E_{1+}$, and $E_{2+}$ of DN GaPN are like the $E_0$, $E_1$, and $E_2$ of GaP, respectively. The $E_{0-}$, $E_{1-}$, and $E_{2-}$ arise from the N-induced band.

Note that $E_{0-}$ allows DN GaPN to behave as a direct-bandgap material, even though it is not a very strong transition. Shtinkov *et al.*[33] has shown that the N-induced band starts with a large contribution from the $s_N$ orbital, and that the contribution from the $s_N$ orbital decreases in the N-induced band and increases in the original host band as the N content increases. By the $sp^3d^5s^*s_N$ model definition, the $s_N$ orbital has no coupling to the $p$ orbitals, which contribute to the upper valence band states, which is consistent with the picture that the localized states couple weakly to the extended states. The band-to-band momentum matrix elements of interest are calculated and plotted against the nitrogen fraction in Figure 7. As the N content increases in GaPN, the N contribution to the lowest conduction band (−) decreases and that to the second lowest conduction band (+) increases as a result of the increasing $s_c - s_N$ coupling, and hence the momentum matrix elements of the "−" transitions increase and those of the "+" transitions decrease, except for $E_{2-}$ and $E_{2+}$, as the $s_N$ orbital does not couple with the $d$ orbitals by definition. One difference between DN GaAsN and DN GaPN is that the N impurity level lies above the conduction band minimum (CBM) of GaAs, but below the CBM of GaP. Therefore, the $E_{0-}$ of DN GaAsN is more like the $E_0$ of GaAs, and the $E_{0+}$ of DN GaAsN is more like the transition between the valence band maximum (VBM) and the localized N state. This causes the momentum matrix elements of $E_{0-}$ and $E_{0+}$ in DN GaAsN to change in the opposite directions to those in DN GaPN as N fraction increases.

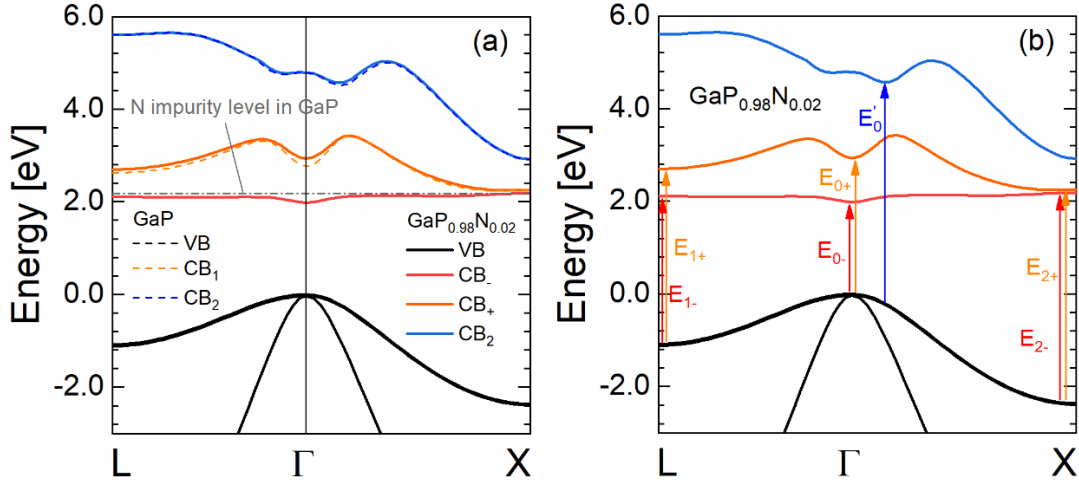

FIG. 6. (a) The highest valence bands (VBs) and lowest conduction bands (CBs) of GaP$_{0.98}$N$_{0.02}$ (solid) and GaP (dashed) calculated using the $sp^3d^5s^*s_N$ and the $sp^3d^5s^*$ tight-binding models, respectively. The N orbital energy in GaP is also indicated by the dashed-dotted line. (b) Important direct transition energies in GaP$_{0.98}$N$_{0.02}$.

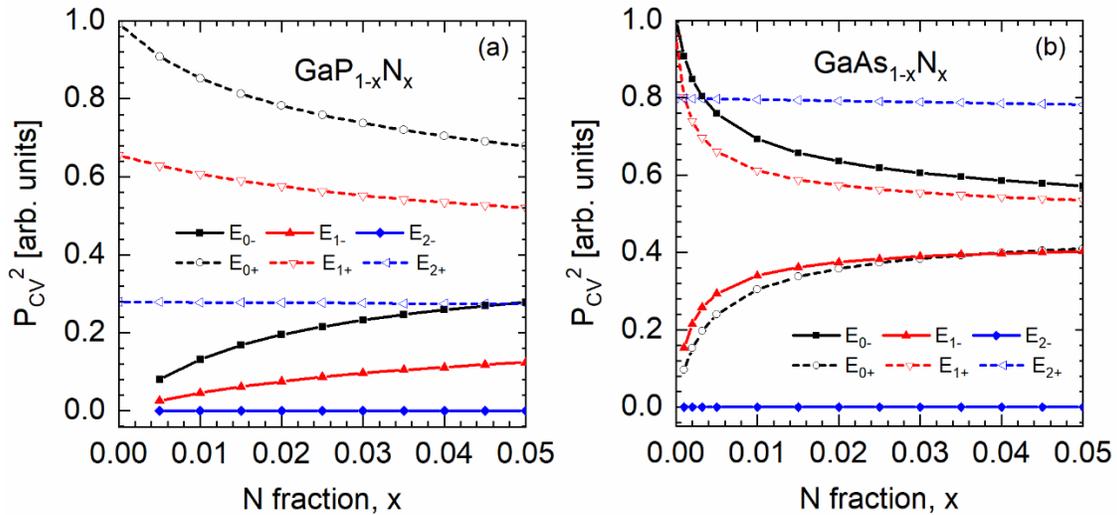

FIG. 7. Band-to-band momentum matrix elements for DN (a) GaPN and (b) GaAsN. At [N]=0, there is no N-induced band or state, therefore no momentum matrix elements are calculated at that point.

To understand the trends of the transitions at the critical points, the $\varepsilon_2$ spectra families calculated for GaPN and GaAsN alloys are plotted in Figure 8. One should note that some of the weaker optical features, such as $E_{0+}$, $E_{1-}$, and $E_{1-} + \Delta_1$, are difficult to see in room-temperature experiments due to lifetime-broadening and indirect transitions, but they are revealed in the calculated results by including only direct transitions. For GaPN alloys with increasing N mole fraction, the fundamental bandgap $E_{0-}$ is pushed downward and the $E_{0+}$ is repelled upward, the $E_{1-}$ transition energy decreases and the $E_{1+}$ energy increases, the $E_0'$ slightly blue shifts while the $E_{2+}$ remains essentially unchanged. No features due to spin-orbit interactions are obvious, because the split-off energies are small in GaP and the N incorporation does not change the split-off energies. These trends agree well with experimental observation[24]. The calculated $\varepsilon_2$ functions for GaAsN show very similar trends. The main difference is that the features due to spin-orbit interaction appear, as the split-off energies for GaAs are much larger than the splitting for GaP, although the features due to $\Delta_0$ are not marked, as they are weak compared to those due to $\Delta_1$. The trends in the calculated $\varepsilon_2$ for GaAsN also agree with reported experiments[21,22].

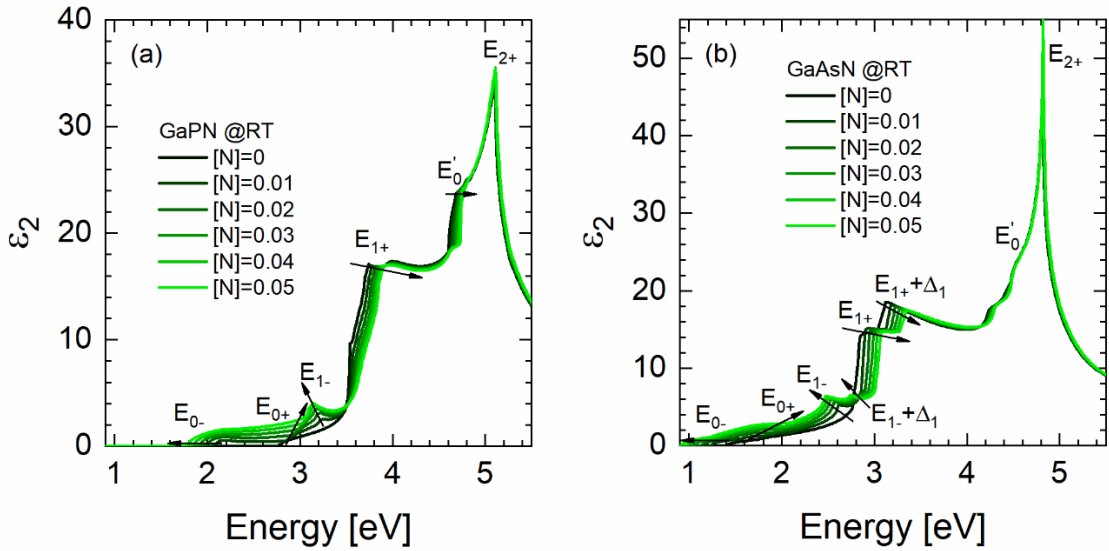

FIG. 8. The calculated imaginary part of the dielectric functions of (a) GaPN and (b) GaAsN alloys. The arrows indicate the trends of the critical transitions with increasing nitrogen composition.

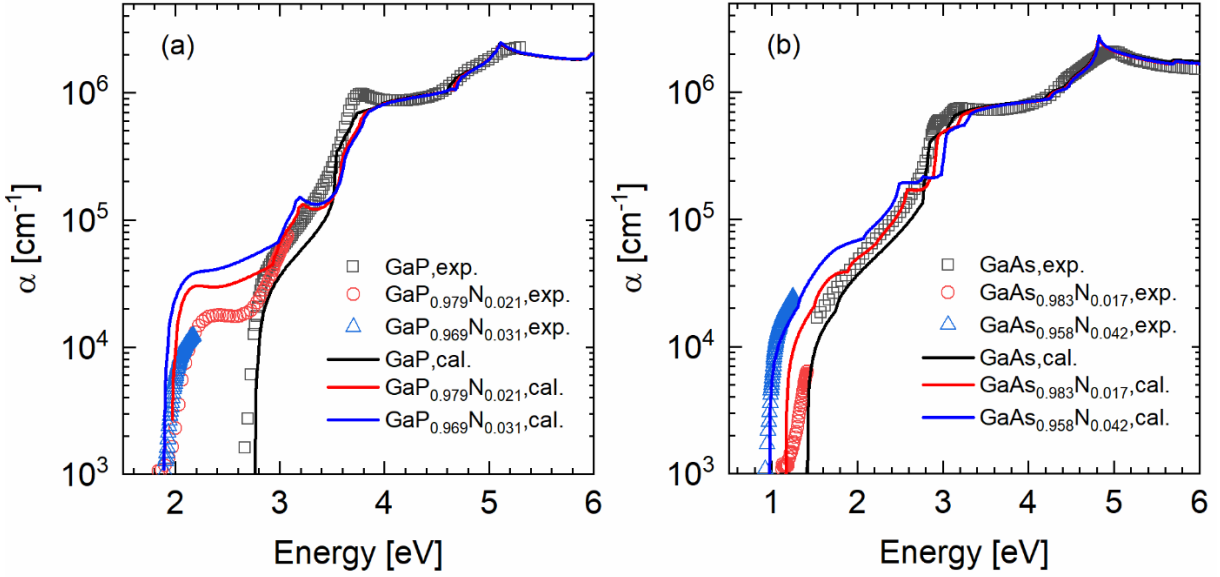

FIG. 9. Measured (exp., solid symbols)[17,20,60,61,66–68] and calculated (cal., solid lines) absorption coefficients of (a) GaP and dilute-N GaPN, (b) GaAs and dilute-N GaAsN.

Figure 9 compares the calculated absorption coefficients with experimental measurements from literature. For GaP, there is no significant absorption below the 2.76 eV, because GaP has an indirect fundamental bandgap of 2.26 eV ($\Gamma - X$), and the smallest direct gap $E_0 = 2.76$ eV at $\Gamma$. However, it is seen, from both experimental and theoretical results, that there is absorption higher than $1 \times 10^4$ cm$^{-1}$ well below 2.26 eV in dilute-N GaPN (due to $E_{0-}$), and that the absorption rises to more than $1 \times 10^5$ cm$^{-1}$ beyond 2.76 eV (due to $E_{0+}$ and $E_{1-}$). Compared to GaP, the rises of the near bandgap absorption of DN GaPN are less steep, as both $E_{0-}$ and $E_{0+}$ in DN GaPN are smaller than $E_0$ in GaP (see Figure 7). As the N fraction increases and the host character increases, the absorption due to $E_{0-}$ becomes stronger, which agrees with experiment[44]. The calculated absorption due to $E_{0-}$ for GaN$_{0.021}$P$_{0.979}$ is about two times stronger than experimental observation. This overestimation may originate from the virtual crystal nature of the $sp^3d^5s^*s_N$ model and thus the inhomogeneity of the dilute nitride alloys are not fully accounted for. Another possible reason could be that the $sp^3d^5s^*s_N$ model uses a single $s_N$ orbital to explain the measured "lumped" bandgap reduction effect caused by N contents, which include isolated N atoms, different N-N pairs, and N clusters. Differentiating different N species' contribution to absorption may correct the overestimation, e.g. by adding different $s_{NN}$ and $s_{Ncluster}$ orbitals. However, the evidence is not strong enough to draw conclusions on this issue. For GaAsN, the $E_{0-}$ state has a majority contribution from the host, giving very similar slopes near the absorption edge to that of GaAs. On the high-energy side, nitrogen has no effect on the absorption, as N incorporation does not impact higher energy bands. Overall, the experimentally measured main effects of the N incorporation into GaP and GaAs are reasonably well described by the $sp^3d^5s^*s_N$ model. Although the $E_{s_N}$ and $s_c s_N \sigma$ parameters for GaPN and GaAsN were determined by fitting to their respective bandgaps only, the above examination of the calculated optical functions suggests that the method used here is suitable for the estimation of the optical properties of dilute nitride alloys.

We checked that the absorption of DN GaPN and GaAsN in the lower energy part ($< 3$ eV) is mostly due to the transition from the valence bands to the lowest conduction band. To obtain more insight into the cause of the increased absorption near the bandgap of these materials, the decomposed imaginary part of the dielectric functions and joint density of states (JDOS) are plotted in Figure 10. The extension of the absorption to lower energies are caused by the redshift of the conduction band edge and hence the extension of the JDOS. However, the increase in the absorption magnitudes, as the N mole fraction increases, cannot be explained by the change in the JDOS, but is mainly a result of the increase in the coupling strength between the valence bands and the lowest conduction bands. Although the coupling strength decreases near Γ in GaAsN and does not change much near $X$ in both DNs (Figure 7), it increases away from Γ and $X$, and overall raises the absorption.

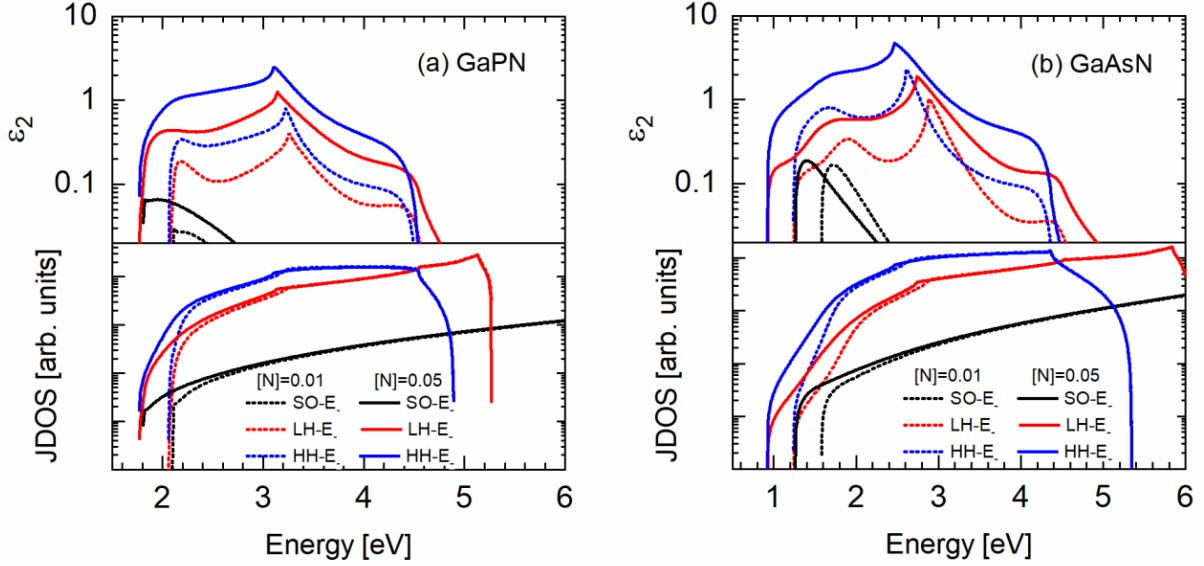

FIG. 10. The imaginary part of the dielectric functions ($\varepsilon_2$) and joint density of states (JDOS) for different interband transitions for GaPN and GaAsN at two N mole fractions, 0.01 (short-dashed) and 0.05 (solid).

## V. OPTICAL CALCULATION FOR DILUTE-N GaPAsN

We construct the $sp^3d^5s^*s_N$ Hamiltonian of DN GaP$_{1-x-y}$As$_y$N$_x$ by linear interpolation (same as the PBL-VCA approach in Appendix A, given as an example for GaPAs) between the $sp^3d^5s^*s_N$ Hamiltonians of DN GaP$_{1-x}$N$_x$ and GaAs$_{1-x}$N$_x$ based on the P/As ratio ($= (1-x-y)/y$). The valence band offset is set to $VBO = VBM(GaAs) - VBM(GaP) = 0.47\ eV$, according to Vurgaftman et al.[62]. Figure 11 plots the calculated bandgaps of DN GaPAsN lattice-matched to (l.m.t.) Si with respect to the N and As fractions. Any vertical line from the bottom axis to the top axis indicates the simultaneous N and As atomic fractions that make a GaPAsN alloy l.m.t. Si. The bandgap of GaP$_{0.816}$A$_{0.134}$N$_{0.050}$ is calculated to be 1.63 eV, about 0.05 eV smaller than that calculated by Almosni and coworkers[12]. The difference lies in that their work was based on $sp^3d^5s^*$ parameters for GaP and GaAs not fit to optical properties, and that they used different $E_s^N$ and $s_cs_N\sigma$ values. The comparison of experimental and calculated bandgaps for DN GaPN and GaAsN are also included in Figure 11. The excellent agreement provides confidence for the predictions of DN GaPAsN alloys.

From the electronic structure, we calculate the optical functions of DN GaPAsN. The calculated absorption coefficients of DN GaPAsN are compared to experimental data as shown in Figure 12. Note that these alloys are not l.m.t. Si, as there is no report of systematic optical measurements for DN GaPAsN l.m.t. Si. Both the theoretical and experimental results follow the same trend as composition varies. The calculated near-bandgap absorption coefficients are higher than the measured values. This is similar to the case as DN GaPN (see Figure 9a), for in both cases, the host materials are indirect bandgap materials, and the $E_{0-}$ states are mostly contributed by the N species. The "dips" in the calculated absorption curves near 3.0–3.5 eV are mostly due to the neglect of lifetime broadening (see Appendix B) and higher-order transitions, which are not the focus of this work.

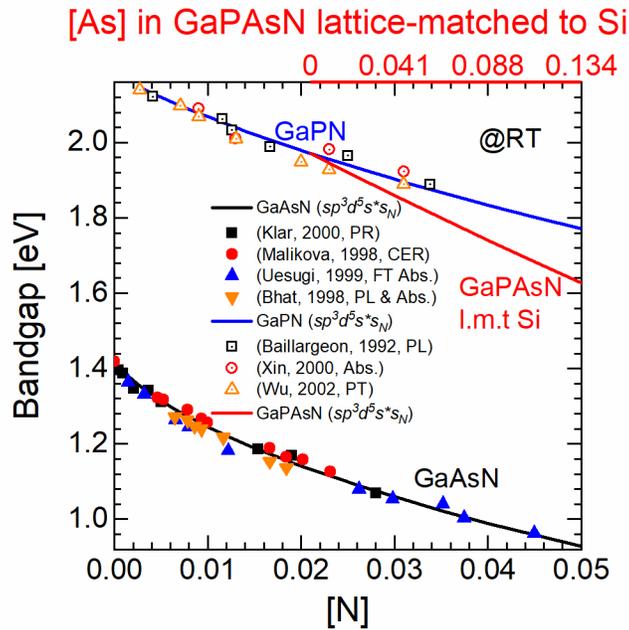

FIG. 11. Calculated (solid lines) and reported experimental[36,39,40,68–71] (symbols) bandgaps of dilute-N GaPN and GaAsN at room temperature (RT). The bandgaps of GaPAsN lattice-matched to Si are plotted from calculation (red solid line). [As] = 0.000 is aligned with [N] = 0.021, that is $GaP_{0.979}N_{0.021}$ is lattice-matched to Si. For [N] < 0.021, there is no GaPN(As) lattice-matched to Si, assuming Vegard's law[72,73]. Calculations carried out here use the $sp^3d^5s^*s_N$ model.

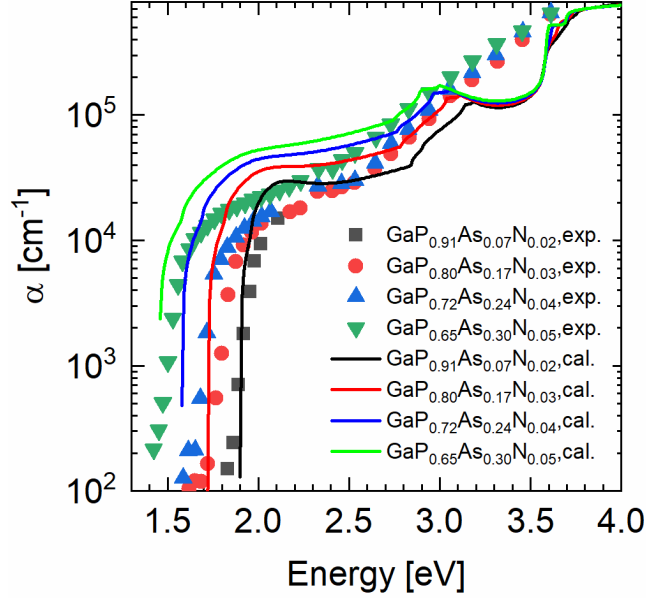

FIG. 12. Absorption coefficients of GaPAsN alloys at room temperature from experiment[18] and from calculation in this work.

To show the effects on absorption of adding N and As into GaP, while keeping the alloy lattice-matched to Si, the calculated absorption coefficients for GaP, and a family of GaPAsN alloys l.m.t. Si are plotted in Figure 13. The addition of N and As largely shifts the absorption edge to lower energy, with $GaP_{0.856}As_{0.101}N_{0.043}$ reaching 1.7 eV (not shown in the plot), an optimal bandgap for series-connected two-junction Si-based tandem solar cells. The interaction between N and the host forms $E_{1-}$ and $E_{1+}$. The former red shifts and the latter blue shifts with increasing N content. The addition of As into GaP causes the $E_1$ transition to red shift and partly compensates the change in the $E_{1+}$ energy due to N. The calculation shows that while the absorption coefficient of the DN GaPAsN alloys in the higher energy range remains mostly the same, it becomes stronger in the lower energy range as the N incorporation increases. This can be explained by the increase in the coupling between the lowest conduction band and the valance bands, similar to the cases of DN GaPN and GaAsN.

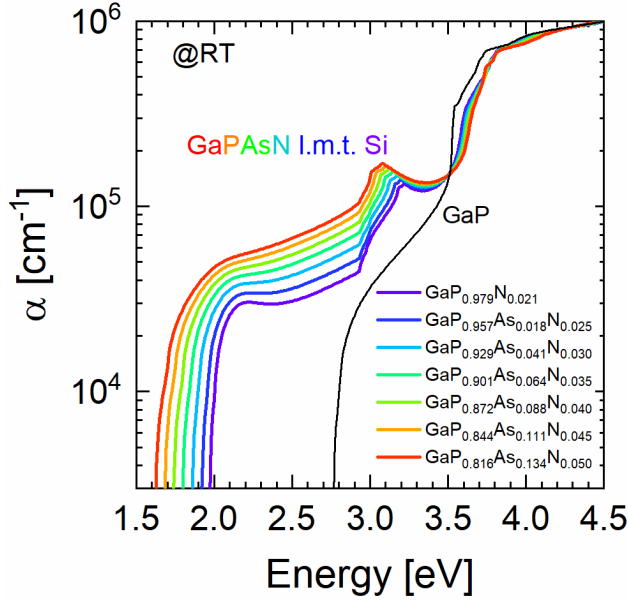

FIG. 13. The absorption coefficients calculated for the family of GaPAsN lattice-matched to Si, and that for GaP shown as a reference.

## V. CONCLUSION

A genetic algorithm is used to optimize the $sp^3d^5s^*$ tight-binding parameters for GaP and GaAs for optical applications. To reach good agreement with optical experiments, it is important to include optical transition energies into the parameter fitting process. The $sp^3d^5s^*s_N$ parameters for DN GaPN and GaAsN are then obtained by fitting to bandgaps without other optical information. This allows reasonable agreement between the calculated and measured optical functions of DN GaPN and GaAsN, and adds to the argument that the $sp^3d^5s^*s_N$ model is an effective one to describe DNs, though more N-related orbitals may be needed to account for the detailed effects of different types of N species. In the end, the bandgaps and optical response functions of DN GaPAsN lattice-matched to Si are calculated. These alloys become stronger absorbers (or emitters) in the lower energy range as the N incorporation increases.

**ACKNOWLEDGMENTS**

This material is based upon work primarily supported by the National Science Foundation (NSF) and the Department of Energy (DOE) under NSF CA No. EEC-1041895. Any opinions, findings and conclusions or recommendations expressed in this material are those of the author(s) and do not necessarily reflect those of NSF or DOE.

**APPENDIX A: CALCULATIONS FOR GaPAs ALLOYS**

For the calculations for GaPAs alloys, we construct the alloy's $sp^3d^5s^*$ Hamiltonian as a linear interpolation of the GaP and GaAs $sp^3d^5s^*$ Hamiltonians, which we refer to as the parent-bond-length virtual crystal approximation (PBL-VCA),

$$H_{AB_xC_{1-x}} = xH_{AB} + (1-x)H_{AC}. \tag{A1}$$

For comparison, we also calculate the TB Hamiltonian for GaP$_x$As$_{1-x}$ through a linear combination of the Hamiltonians of the ending binary semiconductors strained to the alloy ensemble bond length determined by Vegard's law, as done in a previous work[74], which we call the ensemble-bond-length virtual crystal approximation (EBL-VCA),

$$H_{AB_xC_{1-x}} = xH_{AB}(\epsilon_{AB}) + (1-x)H_{AC}(\epsilon_{AC}), \tag{A2}$$

$$\epsilon_X = \frac{d_{ensemble} - d_X}{d_X}. \tag{A3}$$

where $d_X$ is the bond length of material X, and $d_{ensemble}$ is the alloy ensemble bond length. To account for strain, the two-center orbit interaction energies are scaled by the inverse distance power law using the scaling powers from the work by Jancu *et al.*[46]. Although the present Slater-Koster-type TB parameters are different from Jancu *et al.*'s, the scaling of the orbital interactions should not change significantly, hence this method should work as a first approximation.

Figure 14 plots the $E_\Gamma$ and $E_X$ calculated with PBL-VCA and EBL-VCA. The results from PBL-VCA follow closely to the experimental trend[75,76], and the direct-indirect crossover occurs near [P] = 0.46, close to the measured values[62,77]. However, straining the Ga-As and Ga-P bonds to the ensemble bond length overestimates the bowing of $E_X$, and shifts the direct-indirect crossover away from the experimental values. The better performance of PBL-VCA over EBL-VCA is supported by valence force field calculations[74,78]. Bellaiche, Wei, and Zunger found that in the impurity limits of GaAsN, GaPAs alloys, the averaged III-V bond lengths are very close to those in their parent binary semiconductors[78]. Nestoklon, Benchamekh, and Voisin showed that in Ga$_{0.6}$In$_{0.4}$As, the averaged Ga-As and In-As bond lengths are closer to those in bulk GaAs and InAs, respectively, than to the ensemble averaged bond length[74]. For the construction of the $sp^3d^5s^*s_N$ Hamiltonian for GaP$_x$As$_y$N$_{1-x-y}$, we first construction the $sp^3d^5s^*$ Hamiltonian for GaP$_{x/(x+y)}$As$_{y/(x+y)}$, and then account for the nitrogen effects with the $E_{s_N}$ and $s_c s_N \sigma$ parameters.

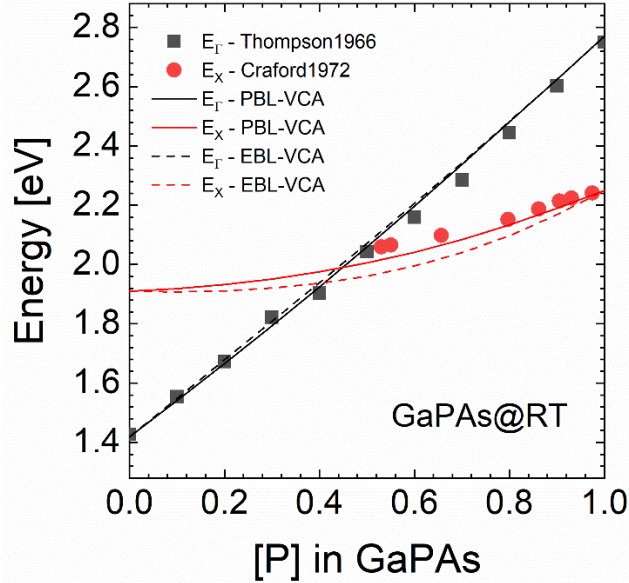

FIG. 14. Comparison of lowest conduction band energies at Γ and X with respect to the valence band maximum. Solid symbols indicate experimental values[75,76], while solid lines and dashed lines are calculated with PBL-VCA and EBL-VCA, respectively.

Continued with the PBL-VCA approach, the refractive indices of some GaPAs alloys are calculated to compare with reported experiment (Figure 15). GaP$_{0.125}$As$_{0.875}$ is a direct-bandgap material, and GaP$_{0.625}$As$_{0.375}$ is indirect. In the lower energy range, the calculated functions agree reasonably with the measured data. The maximum deviation of the calculated values away from the experimental data points is 3.8%. In the higher energy range, the two peaks, due to $E_1$ and $E_1 + \Delta_1$, blue shift as P fraction increases.

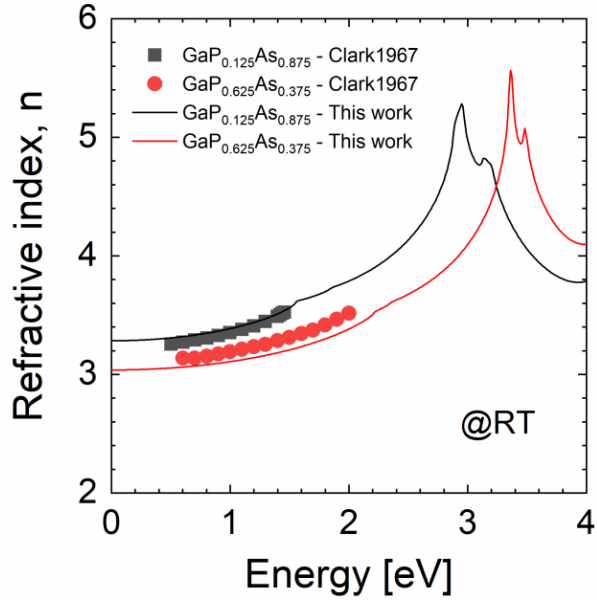

FIG. 15. The real part of the index of refraction of GaPAs alloys from experiments[79] (symbols) and from calculation in this work (curves).

**APPENDIX B: LIFETIME BROADENING EFFECTS ON OPTICAL FUNCTIONS**

The excited states of a system only have finite lifetimes. To account for this finite lifetime effect, one can do the following substitution into (1),

$$\delta(E_c - E_v - \hbar\omega) \to \frac{1}{\pi}\frac{W}{W^2 + (E_c - E_v - \hbar\omega)^2}. \quad (B1)$$

where $W$ represents the broadening in energy.

To see the lifetime broadening effects of the optical response of the dilute nitrides, we applied (B1) for some DN GaAsN alloys, with a constant $W$ for all photon energies for simplicity. As shown in Figure 16a, the most prominent effect is that the broadening reduces the magnitudes of the $E_1$ and $E_2$ peak series. It also fills the "dips" between $E_{1-}$ and $E_{1+}$. These enhance the agreement between the calculations and experiments. A large broadening also smooths out the features at lower energies, and raises the calculated response above the measured values (Figure 16b). In reality, a complicated broadening scheme, with different broadening at different parts of the spectra, should be used. It also indicates that the lifetime broadening alone cannot explain all the discrepancy between the calculation and experiment.

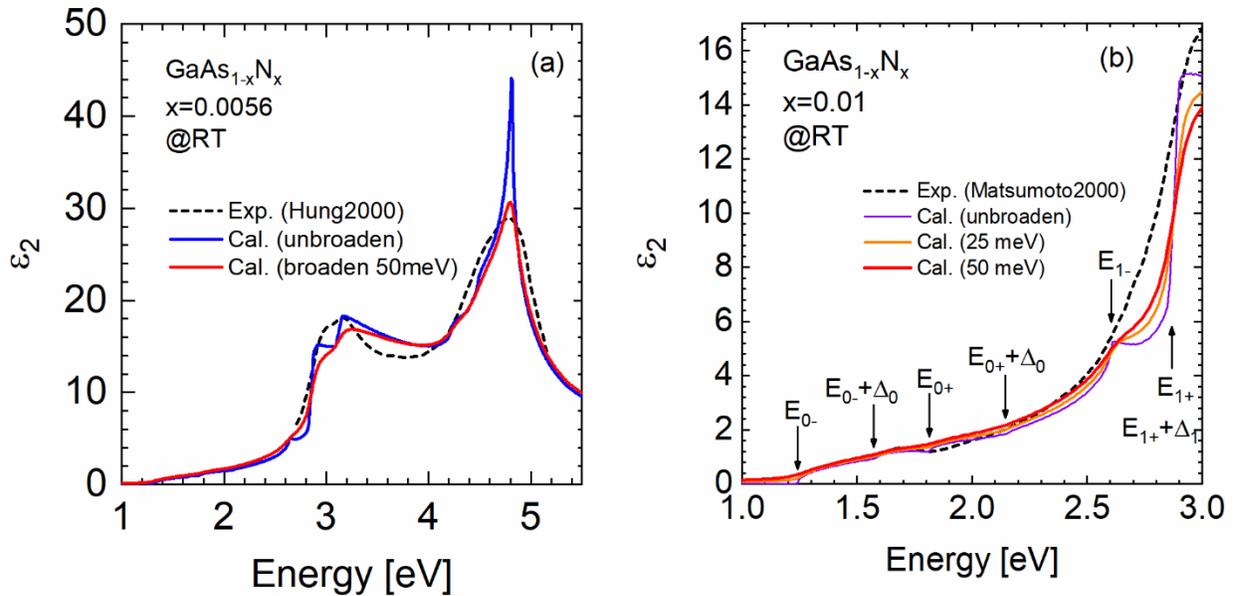

FIG. 16. The imaginary part of the dielectric functions of $GaAs_{1-x}N_x$ at compositions (a) x=0.0056 and (b) x=0.01. The dashed lines are from experiments[22,80], and the solid lines are from calculations.